\documentclass[useAMS]{mn2e}
\usepackage{epsfig}
\usepackage{aasj}
\usepackage{times}

\newcommand{\gtrsim}{\ga}
\newcommand{\lesssim}{\la}
\newcommand{\bi}{\bibitem[]{}}
\newcommand{\epsscale}[1]{}
\newcommand{\plotone}[1]{\begin{center}\epsfig{file=#1,width=8.5cm}\end{center}\vspace{-1cm}}
\newcommand{\plottwo}[2]{\begin{center}\epsfig{file=#1,width=8.5cm}\end{center}\vspace{-1.5cm}\begin{center}\epsfig{file=#2,width=8.5cm}\end{center}\vspace{-1cm}}

\title[Detectability of Extrasolar Planets]{Detectability of Extrasolar Planets in Radial Velocity Surveys}

\author[A. Cumming]{Andrew Cumming\thanks{Hubble Fellow}\\Department of Astronomy and Astrophysics, University of California, Santa Cruz, CA 95064, USA\\email: cumming@ucolick.org}

\begin{document}

\maketitle

\begin{abstract}
Radial velocity surveys are beginning to reach the time baselines required to detect Jupiter analogs, as well as sub-Saturn mass planets in close orbits. Therefore it is important to understand the sensitivity of these surveys at long periods and low amplitudes. In this paper, I derive analytic expressions for the detectability of planets at both short and long periods, for circular and eccentric orbits. In the long period regime, the scaling of the detection threshold with period depends on the desired detection efficiency. The 99\% velocity threshold scales as $K\propto P^2\propto a^3$, whereas the 50\% velocity threshold scales as $K\propto P\propto a^{3/2}$. I suggest an extension of the Lomb-Scargle statistic to Keplerian orbits, and describe how to estimate the false alarm probability associated with a Keplerian fit. I use this Keplerian periodogram to investigate the effect of eccentricity on detectability. At short periods, detectability is reduced for eccentric orbits, mainly due to the sparse sampling of the periastron passage, whereas long period orbits are easier to detect on average if they are eccentric because of the steep velocity gradients near periastron. Fitting Keplerian orbits allows the lost sensitivity at short orbital periods to be recovered for $e\lesssim 0.6$, however there remain significant selection effects against eccentric orbits for $e\gtrsim 0.6$, and the small number of highly eccentric planets discovered so far may reflect this. Finally, I present a Bayesian approach to the periodogram which gives a simple derivation of the probability distributions of noise powers, clarifies why the periodogram is an appropriate way to search for long period signals, and emphasises the equivalence of periodogram and least squares techniques.
\end{abstract}

\begin{keywords}
binaries: spectroscopic --- methods: statistical --- planetary systems
\end{keywords}

\section{Introduction}

Precise radial velocity surveys have made detection of Jupiter mass companions to nearby stars routine, with more than 100 such ``exoplanets'' now known (see Marcy et al.~2003 for a review). As these surveys continue, the accessible parameter space grows towards lower masses and longer orbital periods. For example, recent observations have led to the discovery of Saturn mass planets in close orbits (Fischer et al.~2003), and a population of Jupiters with nearly-circular orbits at distances $\gtrsim 1\ {\rm AU}$ (Vogt et al.~2002; Carter et al.~2003; Jones et al.~2003). In addition, it is now possible to study the statistical occurrence rate and distributions of mass, period, and eccentricity of exoplanets (Vogt et al.~2000; Tabachnik \& Tremaine 2002; Butler et al.~2003; Fischer et al.~2003; Lineweaver \& Grether 2003; Jones et al.~2003; Udry, Mayor, \& Santos 2003), and how these depend on the metallicity of the host stars (Fischer \& Valenti 2003). These distributions contain important information about the planet formation process (Armitage et al.~2002; Ida \& Lin 2004).

This recent work emphasises the need to understand the sensitivity of radial velocity surveys at long periods and low amplitudes. The sensitivity of radial velocity surveys has been discussed previously by several authors, mostly for circular orbits. The Lomb-Scargle periodogram (Lomb 1976; Scargle 1982) is a commonly used technique for searching for periodic sinusoidal signals in unevenly-sampled data, and allows analytic estimates of the detection threshold to be written down for periods less than the duration of the observations (Horne \& Baliunas 1986). This was applied to planet searches by Cochran \& Hatzes (1996) and Nelson \& Angel (1998). Endl et al.~(2002) briefly investigated the detectability of eccentric orbits with the LS periodogram, finding that detectability was significantly reduced for $e\gtrsim 0.5$. The detectability of long period planets, with orbital periods longer than the duration of the observations, was studied by Nelson \& Angel (1998) and Eisner \& Kulkarni (2001). Rather than use the Lomb-Scargle periodogram to measure the significance of the $\chi^2$ fit, Nelson \& Angel (1998) adopted the square of the best fitting amplitude $K^2$. Eisner \& Kulkarni (2001) pointed out that this leads to reduced sensitivity at long periods, because when fitting sinusoids to noise only, one finds that the amplitude is strongly correlated with the fitted phase. They adopted an ``amplitude-phase'' analysis to account for this correlation, and showed that the sensitivity was significantly improved. 

In this paper, I revisit the question of the sensitivity of radial velocity surveys. The main motivations are first to address how to assess the false alarm probability associated with a Keplerian orbit fit to radial velocity measurements, and secondly to derive simple analytic formulae for detection thresholds, including long orbital periods and non-zero orbital eccentricities. I show that significant selection effects operate against highly eccentric orbits, and that at long periods, a seperate analysis of amplitude and phase is not required: the definition of the Lomb-Scargle periodogram in terms of $\Delta\chi^2$ automatically accounts for the correlations between fitted parameters at long periods. In addition, I discuss a Bayesian approach to this problem which will provide a useful basis for future study of the statistical distributions of extrasolar planet properties, as well as emphasising the fundamental equivalence of periodogram and least squares techniques. This equivalence, and the idea of extending the periodogram to non-sinusoidal signals have also been discussed in the literature on Bayesian statistics (see Bretthorst 1988, 2001a,b,c; Scargle 2002; Loredo \& Chernoff 2003).

An outline of the paper is as follows. In \S 2, I review the Lomb-Scargle periodogram, and suggest an extension to Keplerian orbits. In \S 3, I outline a Bayesian approach to the periodogram. In \S 4, I calculate the detection thresholds for short and long periods, and including non-zero eccentricity. The conclusions are presented in \S 5.


\section{Least Squares Fitting and the Periodogram}
\label{sec:stats}

\subsection{Sources of Radial Velocity Variability}

We are interested in detecting the radial velocity wobble due to an orbiting planet given a set of measured radial velocities, observation times, and measurement errors. Often the first indication of the presence of a planet is excess scatter in the radial velocities over the expected amount. A simple way to check for this is to calculate the probability that $\chi^2_{\rm mean}=\sum_j (v_j-\langle v\rangle)^2/\sigma_j^2$ is drawn from a chi-squared distribution (the $\Gamma(\nu/2,1/2)$ distribution; Hoel, Port, \& Stone 1971; Press et al.~1992). Here, $\sigma_i$ is the expected variability for data point $i$, and $\langle v\rangle$ is the mean of the data. A reduced $\chi^2$ much greater than 1 (the exact threshold depending on the desired false alarm probability) indicates excess variability.

However, there is some uncertainty in predicting the expected variability in the radial velocities. Scatter in the radial velocities is expected from statistical and systematic measurement errors, and from intrinsic stellar radial velocity variations, or ``jitter''. The typical measurement error depends on the survey, but is typically $3$--$5\ {\rm m/s}$, and promises to improve towards $\sim 1\ {\rm m/s}$ in the near future (Mayor et al.~2003; Butler et al.~2004). Jitter is thought to arise from a combination of surface convective motions, magnetic activity, and rotation (Saar \& Donahue 1997). The amount of jitter depends on stellar properties such as rotation rate and spectral type, but is typically $3$--$5\ {\rm m/s}$ for chromospherically quiet stars (Saar, Butler, \& Marcy 1998; Santos et al.~2000). Saar et al.~(1998) used data from the Lick survey to find a rough relation $\sigma_V\approx 5\ {\rm m/s}\ (23\ {\rm d}/P_{\rm rot})^{1.1}$ for G and K type stars (Cumming, Marcy, \& Butler 1999, hereafter CMB99).

In this paper, we will generally ignore these uncertainties by assuming that the noise level is unknown, and looking for the best partitioning of the data into Gaussian noise plus a single planet on a circular or Keplerian orbit. However, we will also discuss the case where the noise $\sigma$ can be predicted in advance. As we will show, this issue is only important for small $N$, when the noise level is difficult to accurately determine from the data.
 
\subsection{Circular Orbits: the Lomb-Scargle Periodogram}
\label{sec:LS}

We first discuss the Lomb-Scargle (LS) periodogram (Lomb 1976; Scargle 1982). Since it involves fitting sinusoids to the data, this is particularly appropriate for circular orbits. Given a set of observation times $\left\{t_j\right\}$, velocities $\left\{v_j\right\}$, and measurement errors $\left\{\sigma_j\right\}$, and a trial orbital frequency $\omega=2\pi/P$, we fit the function
\begin{equation}
f_j=A \cos\omega t_j+B\sin\omega t_j+C
\end{equation}
to the data by minimising $\chi^2$, which we write as $\chi^2_{\rm circ}=\sum_j(v_j-f_j)^2/\sigma_j^2$. The number of degrees of freedom is $\nu=N-3$, since there are three parameters ($A,B,C$) in the model. Here, we have extended the original Lomb-Scargle periodogram by allowing the mean to float at each frequency (Walker et al.~1995; Nelson \& Angel 1998; CMB99), rather than subtracting the mean of the data prior to the fit (the importance of this is discussed by CMB99 and Black \& Scargle 1982). 

The goodness of fit is measured by the LS periodogram power $z$, defined as
\begin{equation}\label{eq:zdef}
z(\omega)={\Delta\chi^2/2\over \chi^2_{\rm circ}/\nu},
\end{equation}
where $\Delta\chi^2=\chi^2_{\rm mean}-\chi^2_{\rm circ}$, and $\chi^2_{\rm mean}=\sum_j(v_j-\langle v\rangle)^2/\sigma_j^2$ is the $\chi^2$ of a fit of a constant to the data. The periodogram power $z(\omega)$ measures how much the $\chi^2$ of the fit improves when a sinusoid of frequency $\omega$ is included. As emphasised by Walker et al.~(1995), this is similar to a classical F-test for comparing fits of different models to data (e.g.~Bevington \& Robinson 1992). To search for a periodicity, we evaluate $z(\omega)$ for a range of frequencies, and look for the maximum value $z_0$. 

The significance of the best fit depends on the ``false alarm probability'', or how often a periodogram power as large as the observed power would arise purely due to noise alone. For a single frequency search, the probability that the power at a given frequency exceeds the value $z_0$ is well-determined analytically for Gaussian
noise; it is 
\begin{equation}\label{eq:prob}
{\rm Prob}(z>z_0)=\left(\chi^2_{\rm mean}\over\chi^2_{\rm circ}\right)^{-\nu/2}=\left(1+{2z_0\over\nu}\right)^{-\nu/2},
\end{equation}
which is the cumulative probability arising from the $F_{2,N-3}$ distribution (Schwarzenberg-Czerny 1998; see also CMB99, Appendix B for a summary). This distribution takes the simple form
\begin{equation}\label{eq:prob2}
{\rm Prob}(z>z_0)=\exp\left(-z_0\right)
\end{equation}
for large $N$.

Scargle (1982) defined the periodogram as $z(\omega)=\Delta \chi^2/2$, in which case equation (\ref{eq:prob2}) is valid for all $N$. This definition is appropriate when we know the noise level $\sigma$ in advance, and is attractive because of the very simple probability distribution of equation (\ref{eq:prob2}). However, in general the noise level is not known in advance as we discussed in \S 2.1, and must be estimated from the data (e.g.~Horne \& Baliunas 1986). In this case, the periodogram must be ``normalized'' by an estimate of the noise obtained from the data. In equation (\ref{eq:zdef}), the normalization factor is the $\chi^2$ of the best-fitting sinusoid. The ``normalized'' periodogram acts to partition the data into two pieces, signal plus noise, and determines the best fitting amplitude for each\footnote{In fact, there was a debate in literature over the appropriate way to ``normalize'' the periodogram, whether by the variance of the data (Horne \& Baliunas 1986; Walker et al.~1995), or by the variance of the residuals to the best fit sinusoid (Gilliland \& Baliunas 1987). Schwarzenberg-Czerny (1998) showed that in fact these normalizations are statistically equivalent, and we will return to this issue in \S 2.4.}.

For a search of many frequencies, each ``independent frequency'' must be counted as an individual trial. The false alarm probability is then
\begin{equation}\label{eq:FA}
F=1-\left[1-{\rm Prob}(z>z_0)\right]^M
\end{equation}
where $M$ is the number of independent frequencies, and $z_0$ is the observed power. For small $F$,
\begin{equation}\label{eq:FA2}
F\approx M\ {\rm Prob}(z>z_0)\hspace{1cm}(F\ll 1).
\end{equation}
The detection threshold $z_d$ is the periodogram power corresponding to some (small) value of $F$ (e.g.~$F=0.01$ for a 99\% detection threshold), i.e.~the value of $z$ exceeded due to noise alone in only a small fraction $F$ of trials. An observed power larger than $z_d$ indicates that a signal is likely present.

The remaining task is to determine the number of independent frequencies $M$. Whereas sines and cosines are orthogonal functions for evenly-sampled data, leading to a statistically independent set of frequencies, this is no longer the case for unevenly-sampled data. The seperation between peaks in the periodogram is $1/T$, giving a simple estimate of the number of independent frequencies $M\approx T\Delta f$, where $\Delta f=f_2-f_1$ is the frequency range searched. Often $f_2\gg f_1$, in which case $M\approx f_2T$. 

A better determination of $M$ is to use Monte Carlo simulations, in which data sets of noise only are generated, with velocities either drawn from a Gaussian distribution, or selected with replacement from the residuals about the mean (the so-called ``bootstrapping'' method, e.g.~Press et al.~1992). The fraction of trials for which the maximum periodogram power exceeds the observed value gives the false alarm probability\footnote{For Gaussian noise, the number of trials necessary to calculate $F$ can be reduced by using the analytic form of the distribution given by equation (\ref{eq:FA}). First, using $M$ as a free parameter, fit the analytic distribution to the distribution of noise powers from a small number of trials $N_{\rm trials}$. Then equation (\ref{eq:FA}) with the fitted value of $M$ allows extrapolation to $F\gg (1/N_{\rm trials})$.}.

It is worth emphasising the effect of the uneven sampling on the number of independent frequencies. For evenly-spaced data, the number of independent frequencies is $N/2$, ranging from $1/T$ to the Nyquist frequency, $f_{Ny}=N/2T$. For unevenly-sampled data, Horne \& Baliunas (1986) found that $M\sim N$ for a search up to the Nyquist frequency (see also Press et al.~1992). This agrees with our simple estimate above since $M\approx f_2T\approx N/2$. However, uneven sampling allows frequencies much higher than the Nyquist frequency to be searched (see discussion in Scargle 1982 and Bretthorst 2001a). In general, $M\gg N$, by a factor of $f_2/f_{Ny}$. For example, a set of 30 observations over 7 years has $f_{Ny}\approx 1/(6\ {\rm months})$. A search for periods as short as 2 days then has $M\approx 85N\approx 2500$.

\begin{figure}
\epsscale{0.8}\plottwo{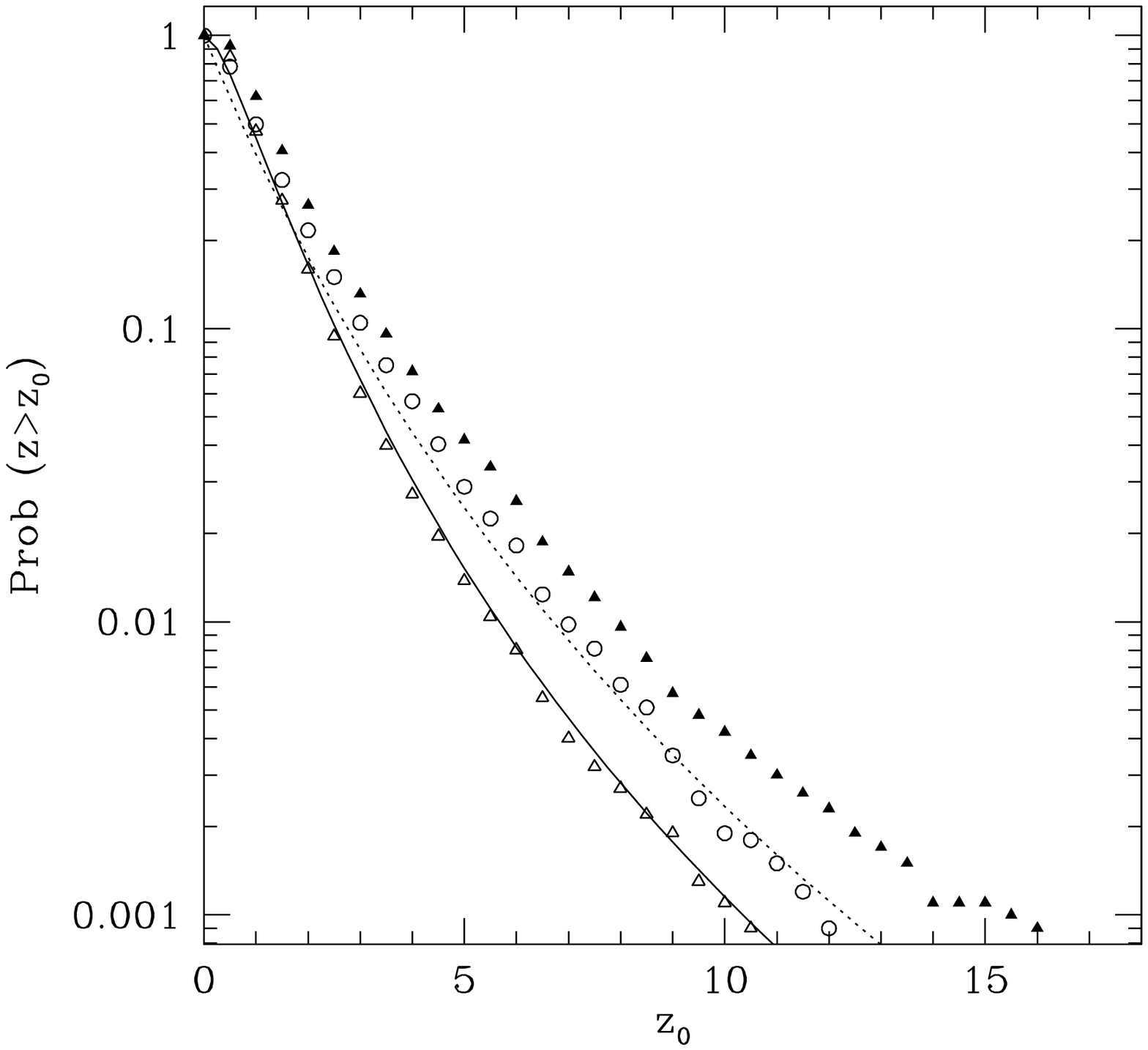}{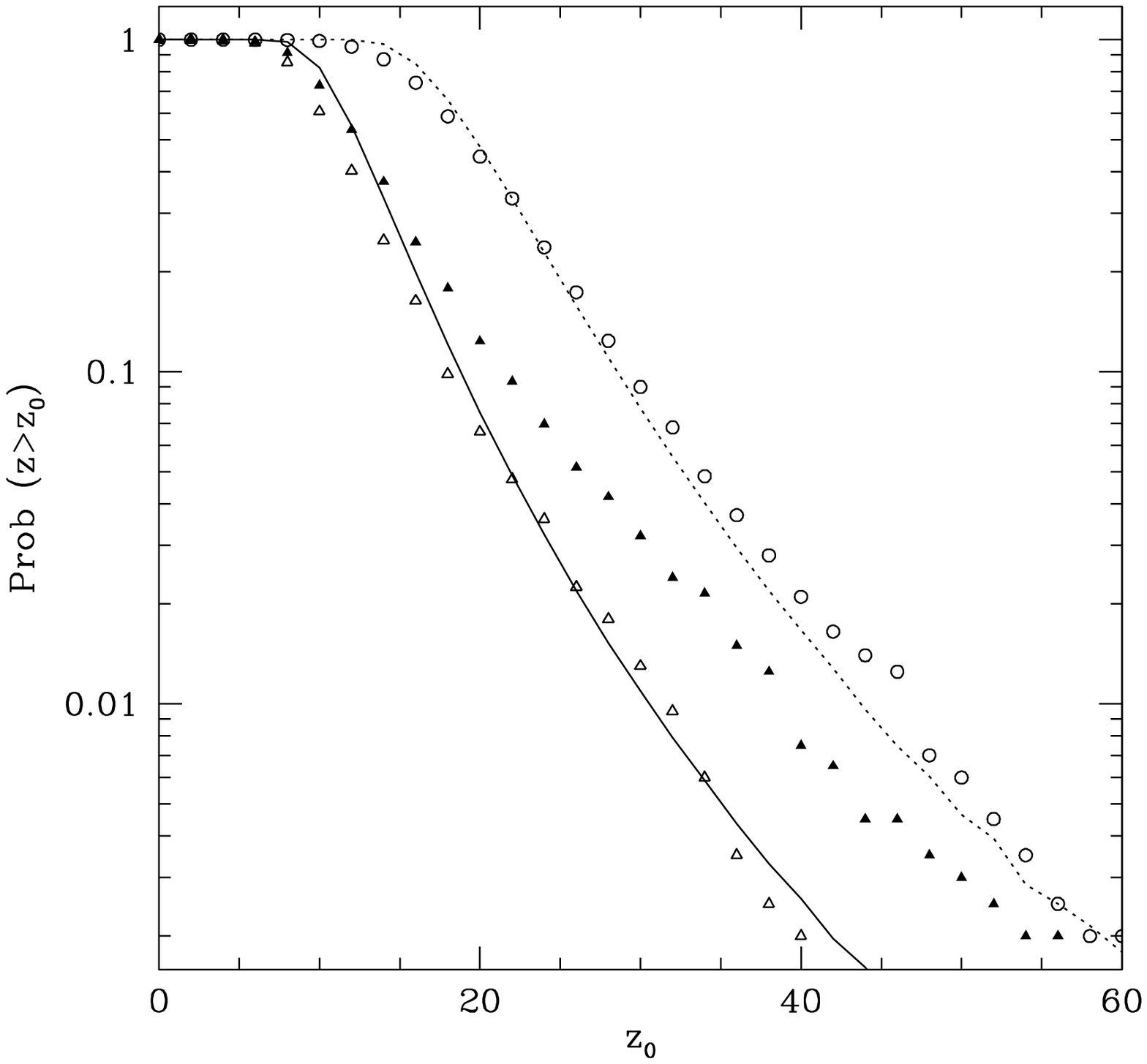}
\caption{Distribution of periodogram powers for Gaussian noise. We take $N=16$ and search at a single frequency (top panel) and for periods in the range 1 day to 10 years (bottom panel). Symbols are the results of numerical simulations with 10000 trials (top panel) or 2000 trials (bottom panel): circles are for the LS periodogram; open triangles are for the Keplerian periodgram, fixing the period during the fit; solid triangles are the Keplerian periodogram with the period allowed to vary during the fit. Dotted and solid curves show the analytic distributions from equations (\ref{eq:prob}) and (\ref{eq:keplerprob}). For the lower panel, the number of independent frequencies needed to fit the numerical results is $M=6000$ for the LS periodogram, and $M=1500$ for the Keplerian periodogram. The duration of the data set is $\approx 1750$ days, giving $M\approx 3T\Delta f$ and $\approx T\Delta f$ respectively.\label{fig:powers}}
\end{figure}

\begin{figure}
\epsscale{1.0}\plotone{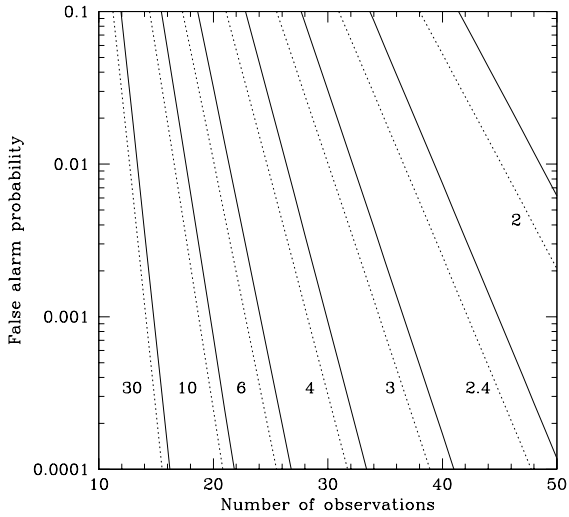}
\caption{The false alarm probability associated with a Keplerian fit to $N$ observations. Each curve shows the FAP from equation (\ref{eq:keplerprob}), labelled by the value of $\chi^2_{\rm mean}/\chi^2_{\rm Kep}$. Solid curves are for $M=3000$, dotted curves are for $M=1000$.\label{fig:fap}}
\end{figure}

\subsection{Generalization of the Periodogram to Keplerian Orbits}
\label{sec:ecc}

The definition of the LS periodogram in equation (\ref{eq:zdef}) suggests an immediate generalization to eccentric orbits (which have 2 extra parameters, the longitude of pericenter $\Omega$, and eccentricity $e$),
\begin{equation}\label{eq:ze}
z_e(\omega)={\Delta\chi^2/4\over \chi^2_{\rm Kep}/\nu},
\end{equation}
where $\chi^2_{\rm Kep}$ is the $\chi^2$ of a fit of a Keplerian orbit to the data (with $\nu=N-5$ degrees of freedom), $\chi^2_{\rm mean}$ the $\chi^2$ of a fit of a constant to the data, and $\Delta\chi^2=\chi^2_{\rm mean}-\chi^2_{\rm Kep}$. Again, the normalization factor in the denominator of equation (\ref{eq:ze}) can be dropped if the noise is known in advance.

The complication in implementing equation (\ref{eq:ze}) is that Keplerian fits are nonlinear, involving relatively slow searches over a complex $\chi^2$ space. Therefore, we use the period and amplitude of the best fitting sinusoid obtained from the LS periodogram as an initial guess for the Keplerian fit. We then use a Levenberg-Marquardt scheme (e.g.~Press et al.~1992) to find the $\chi^2$ minimum, trying several initial starting values for the phases and eccentricity. The minimum $\chi^2$ is then used to calculate $z_e$ from equation (\ref{eq:ze}).

\subsection{The False Alarm Probability of a Keplerian Fit}

Given the redefinition of the periodogram power for Keplerian orbits, the search for significant fits proceeds in a similar way as for circular orbits. For a given data set, we perform a wide frequency search with the LS periodogram, and then use the best fitting sinusoids as starting points for Keplerian fits, calculating the periodogram power from equation (\ref{eq:ze}). 
The significance of the resulting best fitting orbit is determined by a Monte Carlo method in which we make fake data sets containing noise only, and ask how often $z_e$ exceeds the observed value. A similar calculation allows the detection threshold corresponding to a given false alarm probability to be determined.

As in the circular orbit case, there is an analytic estimate for the distribution of noise powers at a single frequency. The distribution of powers for Gaussian noise at a single frequency is given by the $F_{4,N-5}$ distribution (Hoel, Port, \& Stone 1971). Integrating this, we find\footnote{Eq.~(\ref{eq:keplerprob}) becomes ${\rm Prob}(z>z_0)=\left(1+2z_0\right)\exp\left(-2 z_0\right)$ for $N\gg z_0$. Since this is also the cumulative distribution of $\Delta\chi^2/4$, this limit applies to the case where the noise is known in advance, with $z_s$ defined without the normalizing $\chi^2_{\rm Kep}/\nu$ factor.}
\begin{equation}\label{eq:keplerprob}
{\rm Prob}(z>z_0)=\left(1+{\nu+2\over 2}{4z_0\over \nu}\right)\left(1+{4z_0\over\nu}\right)^{-(\nu+2)/2},
\end{equation}
where $\nu=N-5$.

The analytic distribution is compared to numerical calculations in Figure \ref{fig:powers}. The symbols are the results of numerical calculations for data sets with $N=16$. The top panel is for a search at a single frequency, and the bottom panel is for a wide frequency search between periods of 1 day and 10 years. The results for the LS periodogram (circles), and the corresponding analytic distribution (dotted curves) agree extremely well. For the Keplerian periodogram, we show results with the period held fixed during the fit, and with the period allowed to vary during the fit. The solid curves are the analytic distribution of equation (\ref{eq:keplerprob}). The agreement is good for the case where period is held fixed. When the period is allowed to vary during the fit, the powers are systematically larger, since the search algorithm is able to step off the period grid and find a better fit. For this example, the best-fit number of independent frequencies is $M\approx 6000$ for the LS periodogram, and $M\approx 1500$ for the Keplerian periodogram. This is roughly 3 times or equal to the simple estimate $M=T\Delta f$ respectively.

The analytic distribution allows a simple method for estimating the significance of a Keplerian fit to radial velocity data. First, given $\chi^2_{\rm Kep}$ from the Keplerian fit, and $\chi^2_{\rm mean}$ from the fit of a constant to the data, calculate the power $z_0$ using equation (\ref{eq:ze}). An estimate of the false alarm probability is then $F\approx M{\rm Prob}(z>z_0)$ (eqs.~[5] and [6]), where ${\rm Prob}(z>z_0)$ is the probability distribution given by equation (\ref{eq:keplerprob}). The number of independent frequencies is roughly $M\approx T\Delta f$, where $T$ is the duration of the observations, and $\Delta f$ the orbital frequency range searched during the fit.

Figure \ref{fig:fap} shows the FAP as a function of $N$ for different values of $\chi^2_{\rm mean}/\chi^2_{\rm Kep}$. Solid curves are for $M=3000$, dotted curves are for $M=1000$. This Figure can be used for a quick assessment of the FAP associated with a particular Keplerian fit. For example, consider a set of radial velocity measurements with $N=30$ and $\chi^2_{\rm mean}=116$ (or a reduced $\chi^2$ of $4$), and with the best fitting Keplerian having $\chi^2_{\rm Kep}=30$ (reduced $\chi^2$ of $1.2$). Inspection of Figure \ref{fig:fap} shows that in this case, with $N=30$ and $\chi^2_{\rm mean}/\chi^2_{\rm Kep}\approx 4$, the false alarm probability is $\approx 5\times 10^{-4}$.


\section{A Bayesian Approach to the Periodogram}
\label{sec:bayes}

In this section, we show how to understand the Lomb-Scargle periodogram in terms of basic probabilities, i.e.~a Bayesian approach. This leads naturally to the Lomb-Scargle statistic for sinusoid fits, and gives a different way to think about the number of independent frequencies.  A comprehensive and detailed discussion of these issues can be found in a series of papers by Bretthorst (Bretthorst 1988, 2001a,b,c) on the application of Bayesian techniques to spectral analysis, as well as Scargle (2002) and Loredo \& Chernoff (2003). In addition, Ford (2003) discusses the advantages of using likelihood functions to calculate uncertainties in orbital parameters. Those readers interested only in the practical results on detection thresholds should skip ahead to \S \ref{sec:detect}. 

\subsection{The Likelihood Function and Detection Threshold}

We start with the likelihood functions with and without a signal present (Sivia 1996). For a sinusoid fit to the data with parameters $\vec{a}=(K,\phi,P,c)$, the probability of the data given the sinusoid model is
\begin{equation}\label{eq:like}
P(d|K,P,\phi,c)={1\over (\sqrt{2\pi}\sigma)^N}\exp\left(-{\chi^2_{\rm circ}(K,P,\phi,c)\over 2}\right),
\end{equation}
which comes from drawing each observed velocity from a Gaussian distribution. Here, $\chi^2_{\rm circ}$ is not the minimum value of $\chi^2$, but is the value of $\chi^2$ for a particular choice of the parameters $(K,P,\phi,c)$. Since we also have Keplerian fits in mind, we generalize to $m$ parameters, and marginalize or integrate over them to obtain the probability of the data given the presence of a signal,
\begin{equation}\label{eq:pdm}
P(d|1)=\int d^m\vec{a}{1\over (\sqrt{2\pi}\sigma)^N}\exp\left(-{\chi^2_{\rm circ}\over 2}\right).
\end{equation}
Similarly, without a signal present, we may write
\begin{equation}\label{eq:pd0}
P(d|0)=\int dc {1\over (\sqrt{2\pi}\sigma)^N}\exp\left(-{\chi^2_{\rm mean}\over 2}\right).
\end{equation}
where we integrate over the constant term $c$, the only parameter in this case. In these expressions, the notation ``1'' or ``0'' indicates the presence or absence of a planet, and ``$d$'' represents a particular set of radial velocity measurements.

We now follow the detection theory approach of Wainstein \& Zubakov (1962). We write the total probability of the data as
\begin{equation}\label{eq:totalprob}
P(d)=P(1)P(d|1)+P(0)P(d|0),
\end{equation}
where $P(1)$ and $P(0)$ are the prior probabilities that there is or is not a signal present, and we assume here that the only possibilities are that zero or one planet is present in the data. In this framework, the false alarm probability is the probability that there is no signal present given the data, $P(0|d)$. Combining Bayes' theorem, $P(0|d)=P(0)P(d|0)/P(d)$, with equation (\ref{eq:totalprob}), this is
\begin{equation}
F=P(0|d)={1\over 1+\Lambda},
\end{equation}
where we have defined an odds ratio
\begin{equation}
\Lambda={P(1|d)\over P(0|d)}={P(1)P(d|1)\over P(0)P(d|0)}.
\end{equation}
For $\Lambda\gg 1$, the false alarm probability is the inverse of the odds ratio, $F=P(0|d)\approx \Lambda^{-1}$. Therefore, the detection threshold corresponds to a critical value of $\Lambda$. As in the usual frequentist application of the periodogram, the choice of detection threshold represents a compromise between the desired number of false alarms versus false dismissals of real signals (see Appendix of Wainstein \& Zubakov 1962 for a detailed discussion of this issue).

\subsection{Evaluation of the Odds Ratio and Relation to the LS Periodogram}

We now calculate $\Lambda$ and show that it has a direct relationship to the periodogram. The integrals in equations (\ref{eq:pdm}) and (\ref{eq:pd0}) may be evaluated by expanding the integrand around its maximum in terms of the parameters $\vec{a}$. We consider the two cases in which the noise $\sigma$ is known or unknown. In the first case, expanding $\chi^2$ near its minimum gives $\chi^2\approx \chi^2_{\rm min}+\delta \vec{a}\cdot\alpha\cdot\delta\vec{a}$, where the curvature matrix
\begin{equation}
\alpha_{ij}={1\over 2}\left.{\partial^2\chi^2\over\partial a_i\partial a_j}\right|_{\vec{a}=\vec{a_0}}
\end{equation}
is the inverse of the correlation matrix of the $\chi^2$ fit, $C=\alpha^{-1}$ (e.g.~Press et al.~1992). The integral over the parameters is then a standard multidimensional Gaussian integral, giving
\begin{equation}
P(d|1)={1\over \left(\sqrt{2\pi}\right)^\nu\sigma^N}{1\over\sqrt{{\rm det}\,\alpha}}
\exp\left(-{\chi^2_{\rm circ, min}\over 2}\right),
\end{equation}
where $\nu=N-m$. Evaluating $P(d|0)$ in a similar way, and taking the ratio, gives
\begin{equation}\label{eq:lambda1}
\Lambda=\left(2\pi\right)^{\Delta m/2}\ \left[{{{\rm det}\,C_1}\over{{\rm det}\,C_0}}\right]^{1/2}\ {P(1)\over P(0)}\ \exp\left({\Delta\chi^2\over 2}\right)
\end{equation}
where $C_0$ ($C_1$) is the covariance matrix of the fit without (with) a signal present, $\Delta m$ is the number of parameters describing the signal, and $\Delta\chi^2=\chi^2_{{\rm mean},{\rm min}}-\chi^2_{{\rm circ},{\rm min}}$.

When $\sigma$ is unknown, we integrate or marginalize equation (\ref{eq:pdm}) over $\sigma$, taking limits of integration to be $0$ to $\infty$. This gives
\begin{equation}
P(d|1)\propto \int d^m\vec{a}\ \left(\chi^2_{\rm circ}\right)^{-(N/2)},
\end{equation}
with a similar result for equation (\ref{eq:pd0}). The integrals over parameter space are performed as previously, by making a Gaussian approximation near the peak of the integrand. This gives 
\begin{equation}\label{eq:lambda2}
\Lambda=\left(2\pi\right)^{\Delta m/2}\ \left[{{{\rm det}\,C_1}\over{{\rm det}\,C_0}}\right]^{1/2}
\ {P(1)\over P(0)}\ \left({\chi^2_{\rm mean}\over\chi^2_{\rm circ}}\right)^{\nu/2}.
\end{equation}

To see the relation of this result to the periodogram, we write
\begin{equation}
M=\left(2\pi\right)^{-\Delta m/2}\ \left[{{{\rm det}\,C_0}\over{{\rm det}\,C_1}}\right]^{1/2}
\ {P(0)\over P(1)},
\end{equation}
giving
\begin{equation}
F\approx \Lambda^{-1}=M\ {\rm Prob}(z>z_0),
\end{equation}
identical to equation (6), with ${\rm Prob}(z>z_0)$ given by either equation (\ref{eq:prob}) or (\ref{eq:prob2}) depending on whether $\sigma$ is known or unknown\footnote{There is a small difference, which is that $\nu=N-4$ here for a circular orbit rather than $N-3$. The loss of one degree of freedom comes about because of the integration over period. We also mention here that this derivation of the probability distributions applies to either choice of periodogram normalization. Using an analysis of variance approach, Schwarzenberg-Czerny (1998) showed that normalizing the periodogram by the variance of the data or by the variance of the residuals was statistically equivalent, but led to a different form for ${\rm Prob}(z>z_0)$ in each case (see Table 9 in Cumming et al.~1999). In fact, both these distributions are a direct rewrite of the last term in equation (\ref{eq:lambda2}), using the respective definition of $z$.}.

For sinusoid fits, we have recovered the LS periodogram, but with a new interpretation of $M$. For independent parameters, the covariance matrix is diagonal, so that $\det C\approx \prod_i \delta a_i$, where $\delta a_i$ is the uncertainty in parameter $i$ (e.g.~Press et al.~1992). For circular orbits, this gives an estimate
\begin{equation}
M\approx {\Delta P\over \delta P}{\Delta K\over\delta K\delta \phi}
{P(0)\over P(1)}.
\end{equation}
The first term is the estimate for $M$ that we gave in \S 2.2, the range of frequencies searched divided by the frequency resolution. However, we see that in a Bayesian approach, $M$ also includes the range of amplitudes and phase considered, and the prior probability of a signal being present. This corresponds to a different picture in which rather than making many trials searching for the lowest $\chi^2$, we have instead integrated over all possible values of the parameters, weighting each choice by its relative probability. For fits of Keplerian orbits, the probability distribution of $z_e$ is given by equation (\ref{eq:keplerprob}), which is similar to but slightly different from equations (\ref{eq:lambda1}) and (\ref{eq:lambda2}). The significance of this difference is not clear.

The most useful application of this method is as a powerful tool for population analyses. Here, we have discussed the choice between detection and non-detection so that the connection to the periodogram could be seen. However, the advantage of the Bayesian approach is that no decision regarding detection needs to be made. Instead, both possibilities can be included, with the probability of each calculated as $P(1|d)$ and $P(0|d)$. For example, integrating equation (\ref{eq:like}) over the ``nuisance'' parameters $\phi$ and $c$ gives ${\rm Prob}(P,K|d)$ for each star. This can be used to study the underlying mass and orbital period distribution. Further integration of ${\rm Prob}(P,K|d)$ gives ${\rm Prob}(P|d)$, equivalent to the periodogram, or ${\rm Prob}(K|d)$ from which limits on $K$ can be derived. Extension of this approach to eccentric orbits, or multiple planets, is straightforward in principle, but in practice involves more complex integrals (see Ford 2003 for a suggestion of how to evaluate them).


\section{Velocity Thresholds}\label{sec:detect}

\begin{figure}
\epsscale{1.2}\plotone{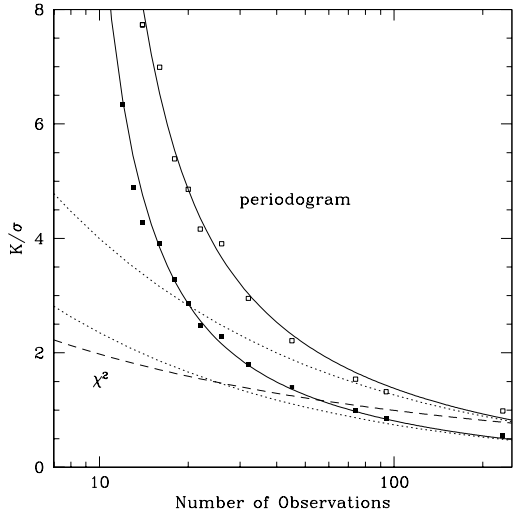}
\caption{Signal to noise ratio $K/\sigma$ that can be detected with $N$ observations, and $99$\% (upper curve) and $50$\% detection efficiency (lower curve). We assume short period ($P<T$) circular orbits, and use the LS periodogram. The solid lines show the analytic result for $M/F=10^6$. The points show numerical calculations for data sets with the indicated values of $N$, and a realistic spacing of observation times. The dotted lines show the result when the noise level is known in advance. The dashed curve shows the 50\% detection efficiency curve for a $\chi^2$ test for excess variability.\label{fig:signoise}}
\end{figure}

\begin{figure}
\epsscale{1.0}\plottwo{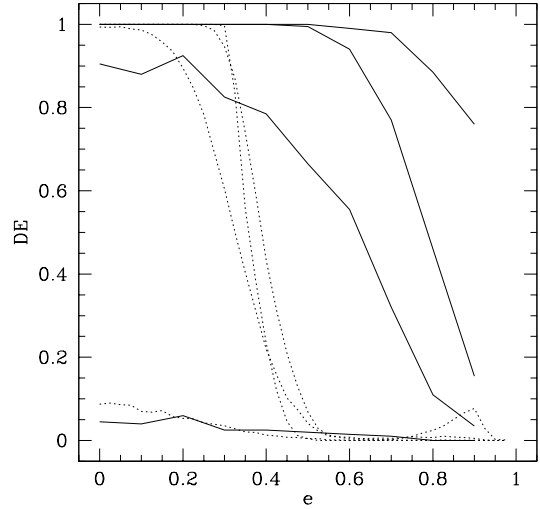}{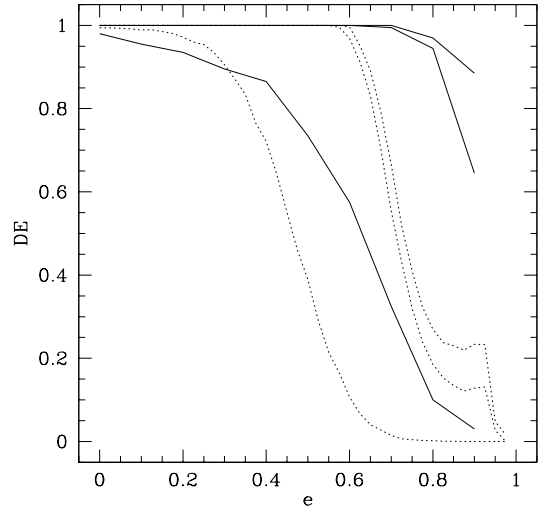}
\caption{Detection efficiency (DE) as a function of eccentricity for a time series with $N=16$ (top panel) and $N=39$ (bottom panel). We use a period $P=100$ days, and show signal to noise ratios $K/\sigma=2,5,10$, and $\infty$ (top panel) and $K/\sigma=2,10$, and $\infty$ (bottom panel). Dotted lines show the results using the LS Periodogram (fitting sinusoids); solid lines use Keplerian fits.\label{fig:detect_eff}}
\end{figure}

In this section, we use the LS periodogram and the Keplerian periodogram discussed in \S 2 to derive analytic expressions for the velocity thresholds. We discuss orbital periods $P$ shorter and longer than the duration of the observations $T$ separately.

\subsection{Short Periods ($P<T$)}
\label{sec:short}

We first consider circular orbits, and use the LS periodogram. Figure \ref{fig:signoise} shows the signal to noise ratio $K/\sigma$ required for 50\% and 99\% detection probability as a function of the number of observations. The squares show numerical calculations for sets of $N$ observations with realistic spacing between observation times. For each set of observation times, we use a Monte Carlo method to determine the detection threshold, generating sequences of velocities drawn from a Gaussian distribution, and finding the power exceeded in 0.1\% of trials (the 99.9\% detection threshold). We search for periods between 2 days and 10 years. We then make fake data sets with increasing signal amplitude $K$ until the signal is detected in either 50\% or 99\% of trials. 

\begin{figure}
\vspace{-2cm}
\epsscale{1.0}\plotone{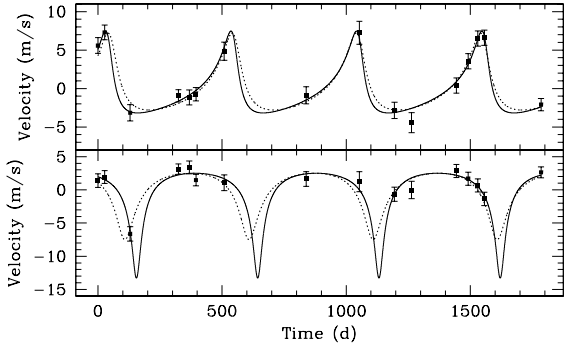}
\caption{Examples of velocity curves with $e=0.5$ that are (top panel) and are not (bottom panel) detected. The dotted line in each case shows the true orbit; the points are the observed velocities; and the solid curve shows the best fitting orbit. In both cases, the solid curve gives a lower $\chi^2$ than the dotted curve. The lower panel has only a single measurement during the periastron passage, and is not a significant detection.\label{fig:detect_example}}
\end{figure}

\begin{figure}
\epsscale{1.0}\plotone{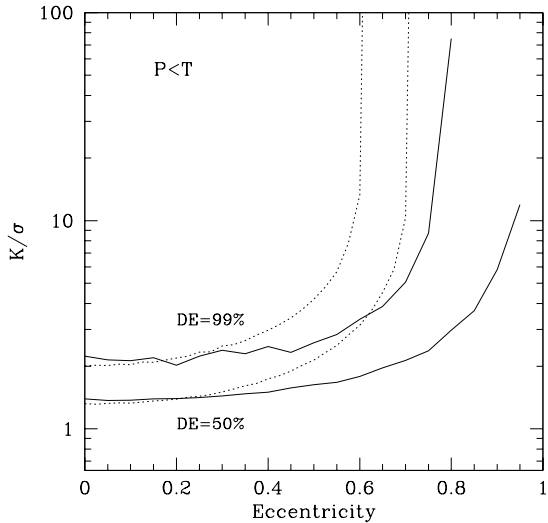}
\caption{The effect of eccentricity on the velocity threshold for $N=39$. The dotted curves are for the LS periodogram; the solid curves are for Keplerian fits. The rapid increase in the 99\% detection efficiency (solid) curves at $e\approx 0.8$ is due to the fact that very eccentric orbits are not always detected even when the signal to noise is very large.\label{fig:e1}}
\end{figure}

To obtain an analytic estimate, we first write down the detection threshold $z_d$ corresponding to a critical false alarm probability $F$ (where $F\ll 1$). Equations (\ref{eq:prob}) and (\ref{eq:FA2}) give
\begin{equation}
z_d={\nu\over 2}\left[\left({M\over F}\right)^{2/\nu}-1\right],
\end{equation}
for a given number of independent frequencies $M$. When a signal of amplitude $K$ is present, the periodogram power $z_s$ has a distribution of values because of noise fluctuations. For large signal amplitude or $N$, the distribution of periodogram powers when a signal is present is Gaussian with mean $\langle z_s\rangle$, and variance $2\langle z_s\rangle$ (Groth 1975), where $\langle z_s\rangle=(\nu/2)(K^2/2\sigma^2)$ (Scargle 1982; Horne \& Baliunas 1986). In this limit, the probability of detecting a signal with mean amplitude $\langle z_s\rangle$ for a given detection threshold $z_d$ is
\begin{equation}\label{eq:erf}
P_\mathrm{detect}(\langle z_s\rangle; z_d)\approx {1\over 2}\left[1+\mathrm{erf}\left({\langle z_s\rangle-z_d\over 2\sqrt{\langle z_s\rangle}}\right)\right].
\end{equation}
Setting $z_d=\langle z_s\rangle$ gives the signal to noise ratio needed to detect the signal 50\% of the time,
\begin{equation}\label{eq:sn1}
{K_0\over\sqrt{2}\sigma}=\left[\left({M\over
F}\right)^{2/\nu}-1\right]^{1/2},
\end{equation} 
or, for large $N$,
\begin{equation}\label{eq:sn}
K_0={\sigma\over\sqrt{N}}\left[4\ln\left({M\over F}\right)\right]^{1/2}\hspace{0.5
cm}(N\gg 1),
\end{equation}
which shows the expected $1/\sqrt{N}$ behavior\footnote{To derive this limit, write $(M/F)^{2/\nu}=1+x$, where $x\ll 1$.}. The solid lines in Figure \ref{fig:signoise} show this analytic estimate for $M/F=10^6$. We include a multiplicative factor of $1.7$ for the 99\% detection probability curve (this factor is given by equation (\ref{eq:erf}) with $P_{\rm detect}=99$\%).

Figure \ref{fig:signoise} shows that $N\gtrsim 10$--$20$ is required to be able to detect an orbit with $K\approx 2$--$4\sigma$, with $N\gtrsim 50$ required to reach amplitudes as small as $K\sim\sigma$. How does that compare to the case where we know the noise level $\sigma$? The appropriate formula is then equation (\ref{eq:sn}) for all $N$. The dotted curves in Figure \ref{fig:signoise} show the 50\% and 99\% detection probability curves in this case. Knowing the noise level in advance gives a significant improvement, allowing a detection of the signal for $N<10$. The sparse sampling of the data may increase the detection threshold somewhat over this estimate (Nelson \& Angel 1998). It is also interesting here to compare the detection threshold of the LS periodogram with a $\chi^2$ test for variability. The dashed line in Figure \ref{fig:signoise} shows the signal to noise ratio that gives a $\chi^2$ exceeding the expected value 50\% of the time, assuming that the noise level $\sigma$ is a known quantity. Here we choose the detection threshold to correspond to a FAP of 0.1\%. Excess variability at the $\approx 2\sigma$ level is apparent with only a handful of observations.

We next discuss the effect of non-zero eccentricity. Figure \ref{fig:detect_eff} shows the detection probability as a function of eccentricity for signal to noise ratios of $K/\sigma=2,5,10$ and $\infty$, where $\infty$ means that we sample the velocity curve without adding noise. The top panel shows results for a data set with 16 observations, the lower panel for 39 observations. The dotted curves show the detection probability using the LS periodogram. The solid curves show the detection probability using the Keplerian periodogram defined in equation (\ref{eq:ze}). Not surprisingly, the LS periodogram fails to detect orbits with large eccentricities, particularly for small $N$, whereas fitting Keplerian orbits increases the detection probability for eccentric orbits.

\begin{figure}
\epsscale{1.2}\plotone{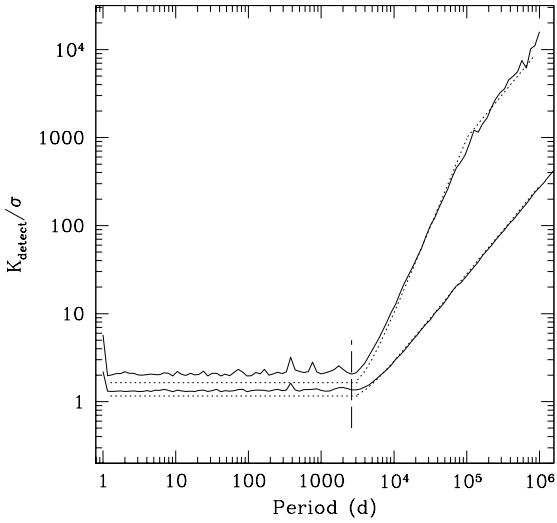}
\caption{Velocity threshold for a star with $N=39$, and detection efficiency 50\% (lower curve) and 99\% (upper curve). The vertical dashed line shows the duration of the observations ($\approx 2600$ days). The detection threshold is set at $99$\% significance for a search down to 1 day period. The dotted lines show the analytic result for $M=5000$ and $F=0.01$. The 50\% curve scales as $K\propto P$ for $P>T$; the 99\% curve scales as $K\propto P^2$ for $P<(\pi/8F_d)T\approx 40T$, and $K\propto P$ thereafter.\label{fig:threshold}}
\end{figure}

\begin{figure}
\epsscale{1.2}\plotone{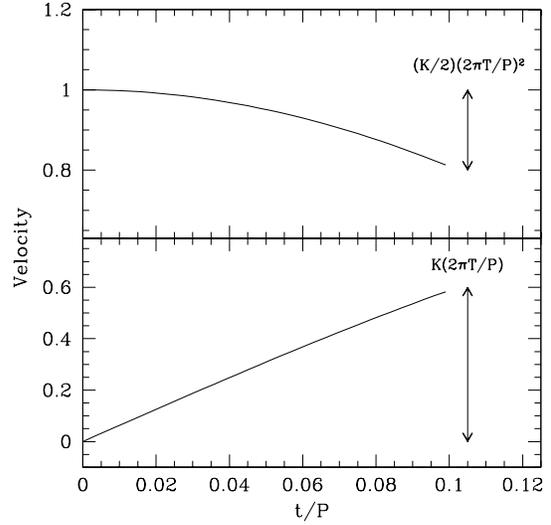}
\caption{Example of "cosine-like" and "sine-like" velocity variations for observations covering $1/10$ of an orbital period . For the cosine case (upper panel), the variation in radial velocity is $\Delta v\approx (K/2)(2\pi T/P)^2$; for the sine case (lower panel), the variation is $\Delta v\approx K(2\pi T/P)$. If $K_0$ is the velocity amplitude that can be detected at short periods, the detection threshold in each case is given by $\Delta v=2K_0$ (see text).\label{fig:longperiod}}
\end{figure}

Figure \ref{fig:detect_eff} shows that even when fitting Keplerian orbits, detectability decreases for $e\gtrsim 0.5$. There are two reasons for this. First, even for $K/\sigma\gg 1$, we find detection efficiencies $<1$ ($\approx 80$--$90$\%) for $e=0.9$. In the cases that are not detected, the Keplerian fitting routine fails to find the correct solution. This emphasises the difficulty of finding the global minimum in the complicated $\chi^2$ space for these nonlinear solutions. The second effect, dominating at lower $K/\sigma$, is the uneven sampling of the data, which can lead to a poorly-resolved periastron passage. An example is shown in Figure \ref{fig:detect_example}, in which we show two sets of observations of an orbit with $e=0.5$. Both fitted lightcurves (solid curves) have a lower $\chi^2$ than the true solution (dotted curve), but whereas the data in the upper panel lead to a detection, the data in the lower panel do not. In the lower panel, only a single measurement has been made during the periastron passage. This greatly reduces the $\Delta \chi^2$ when the Keplerian orbit is included in the fit. An additional danger is that a single discrepant data point might arise due to a systematic error, perhaps making the fit in the lower panel of Figure \ref{fig:detect_example} worrying in a real life example.

Figure \ref{fig:e1} summarizes the effect of eccentricity. We plot the signal to noise ratio needed for a detection efficiency (DE) of 50\% or 99\% as a function of eccentricity. The solid curves are for the Keplerian periodogram, and the dotted curves are for the LS periodogram. The rapid increase in $K/\sigma$ for the 99\% DE solid curve at $e\approx 0.7$ is due to the failure to detect even high signal to noise orbits for $e\sim 1$. For $e\lesssim 0.6$, the effect of eccentricity on the amplitude needed for detection is small.

We have not yet discussed the dependence of the velocity threshold on orbital period. Figure \ref{fig:threshold} shows the amplitude needed for 50\% and 99\% DE for $N=39$ as a function of period. We discuss the long period behavior ($P>T$) in the next section. The dotted lines are the analytic result for $M/F=10^6$, and compare well with the numerical calculations. This Figure shows that the amplitude threshold is not very sensitive to period for $P<T$. There is some loss of sensitivity at periods related to a month and a year, and particularly at $P=1$ day, introduced by the time sampling of the data, but this is a small effect even for $N\sim 10$. This insensitivity can also be seen in the curves of Walker et al.~(1995), Nelson \& Angel (1998), and CMB99. It is a result of the uneven sampling which gives good phase coverage for many frequencies (Press et al.~1992). Therefore the velocity threshold $K_0$ characterizes the short period detectability for most periods $<T$.

\begin{figure}
\epsscale{1.2}\plotone{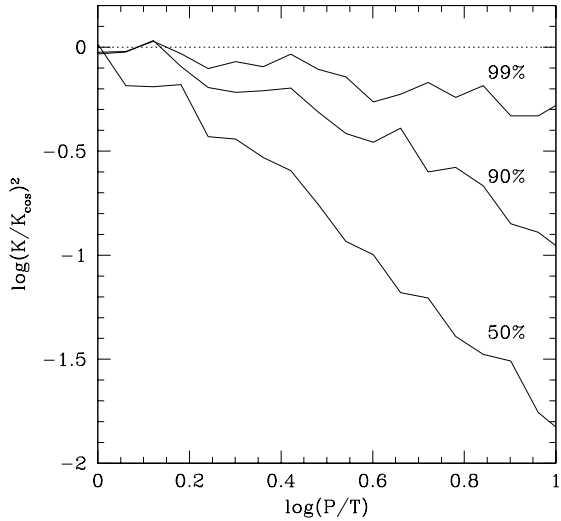}
\caption{Ratio of the 50\%, 90\%, and 99\% velocity-squared thresholds to the cosine like scaling $K^2=4K_0^2/(1-\cos(\pi T/2P))^2$. Comparison of this plot with Figures 8, 9, and 10 of Eisner \& Kulkarni (2001) shows that the LS periodogram and amplitude-phase analysis have equal sensitivities for long periods. The strong deviation for the 50\% curve is because this curve follows a sine-like scaling (eq.~[\ref{eq:long1}]) rather than cosine-like (eq.~[\ref{eq:long2}]).
\label{fig:detect_ratio}}
\end{figure}

\subsection{Long Periods ($P>T$)}
\label{sec:long}

At long periods, the detection sensitivity drops because the observations cover only part of an orbit. For circular orbits, Eisner \& Kulkarni (2001) pointed out that there is a correlation between the fitted phase and amplitude at long periods, and derived analytic expressions for the velocity threshold. Here, we adopt a slightly different approach. It is useful to think of ``sine-like'' and ``cosine-like'' phases, depending on whether the orbit is close to a velocity maximum/minimum or to a zero crossing. These two cases are illustrated by Figure \ref{fig:longperiod}. For $T\ll P$, the velocity variation due to the signal is $\Delta v=K\sin(2\pi T/P)\approx K(2\pi T/P)$ for a sine-like variation, or $\Delta v=K\cos(2\pi T/P)\approx (K/2)(2\pi T/P)^2$ for a cosine-like variation. Setting $\Delta v=2K_0$ then gives an estimate for the velocity amplitude needed for detection.

Averaging over phase introduces slightly different numerical factors. We find that a good approximation to the velocity threshold is
\begin{equation}\label{eq:long1}
K={K_0\over \sin\left(\pi T/2P\right)}
\approx K_0\left({2P\over \pi T}\right)\hspace{1cm}P>T
\end{equation}
for $\epsilon_D<3/4$, or
\begin{equation}\label{eq:long2}
K=\cases{{2K_0\over 1-\cos\left(\pi T/2P\right)}\approx K_0\left({2P\over \pi T}\right)^2 & $T<P<{\pi T\over 8(1-\epsilon_D)}$\cr
{K_0\over 2(1-\epsilon_D)}\left({P\over\pi T}\right) & $P>{\pi T\over 8(1-\epsilon_D)}$\cr}
\end{equation}
for $\epsilon_D>3/4$, where $\epsilon_D$ is the detection efficiency. The scaling $K\propto P$ corresponds to $M_p\propto a^2$; the scaling $K\propto P^2$ corresponds to $M_p\propto a^{7/2}$.

The reason that the scalings depend on the detection efficiency is that a sine-like phase gives a larger $\Delta v$ than a cosine-like phase. So for a 50\% detection threshold ($\epsilon_D=0.5$) the amplitude must be large enough that sine-like phases are detected, but cosine-like phases do not have to be. For a 99\% threshold however ($\epsilon_D=0.99$), almost all phases must be detected, requiring a large amplitude. Eventually, at very long periods, almost all phases are sine-like, and the scaling changes to $\propto P$ once more. Figure \ref{fig:threshold} compares the analytic and numerical results at long periods. For this particular example, the change in scaling of the 99\% threshold from $\propto P^2$ to $\propto P$ can be seen at $P\approx 10^5\ {\rm days}$.

\begin{figure}
\epsscale{1.0}\plotone{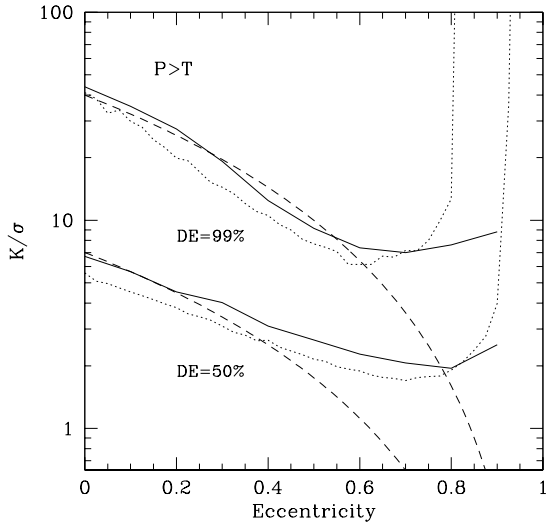}
\caption{For the same data set as Fig.~\ref{fig:threshold}, and an orbital period of $50$ years, the effect of eccentricity on the signal to noise needed for detection 50\% (lower curves) or 99\% (upper curves) of the time. The solid curves are for the Keplerian periodogram; the dotted curves are for the LS periodogram. The dashed curves show an analytic $(1-e)^2$ scaling.\label{fig:longecc1}}
\end{figure}

\begin{figure}
\epsscale{1.2}\plotone{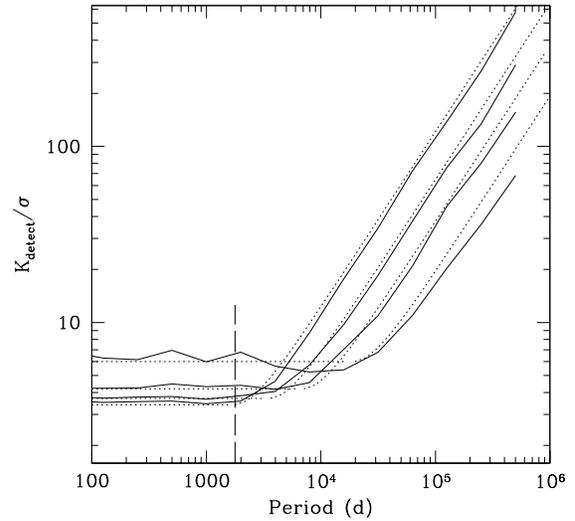}
\caption{For data sets with $N=16$, the effect of eccentricity on the 50\% detection curves. We take $e=0, 0.3, 0.5, 0.7$ (top to bottom at short periods). The dotted curves show the analytic estimates. The vertical dashed curve indicates the duration of the observations.\label{fig:longecc2}}
\end{figure}

In their investigation of detectabilities, Nelson \& Angel (1998) adopted the square of the best fitting amplitude $K^2$ as their test statistic. Eisner \& Kulkarni (2001) pointed out that this leads to reduced sensitivity at long periods, because when fitting sinusoids to noise only, one finds that the amplitude is strongly correlated with the fitted phase. The cosine-like phases have much larger fitted amplitudes than the sine-like phases, and so the velocity threshold has a $K\propto P^2$ scaling. Indeed, our results nicely explain the empirical scaling found by Nelson \& Angel (1998), $K\propto (P/\beta T)^\alpha$, with $\alpha=1.86$ and $\beta\approx 1.45$. Our formula gives $\alpha=2$ and $\beta=\pi/2=1.57$. Eisner \& Kulkarni (2001) adopted an ``amplitude-phase'' analysis to account for this correlation, in which the detection threshold is set by an ellipse in the $K$--$\phi$ plane. They showed that the sensitivity was significantly improved. However, their proposed method is unnecessarily complicated. The correlation between amplitude and phase is automatically included in the LS periodogram, which is much simpler to use since it is defined in terms of the single variable $\Delta\chi^2$. We have implemented the method of Eisner \& Kulkarni (2001) and find that it matches the sensitivity of the periodogram at all periods. Figure \ref{fig:detect_ratio} shows the ratio of the velocity thresholds from the periodogram to the cosine-like scaling of equation (\ref{eq:long2}). Comparison with Figures 8, 9, and 10 of Eisner \& Kulkarni (2001) shows that the LS periodogram has equal sensitivity compared to the amplitude-phase technique.

Walker et al.~(1995) looked for long term periodicities by fitting a quadratic to the data $v=a+b\,t + c\,t^2$, and checking for a significant reduction in $\chi^2$ using an F-test. CMB99 adopted a similar approach but with linear fits to the data. How is this related to the long period sensitivity of the periodogram? In fact, for periods $P\gtrsim 2\pi T$, the periodogram is no longer sensitive to the period. A quadratic fit gives a relation between the amplitude and period
\begin{equation}\label{eq:quadK}
K={P\over 2\pi}\left(b^2+P^2{c^2\over \pi^2}\right)^{1/2},
\end{equation}
so that the best fitting velocity amplitude is determined for all periods ($>T$) by the fit. We have checked the sensitivity to long period circular orbits of an F-test based on quadratic fits to the data, and find that quadratic fits reproduce the sensitivity of the periodogram for long periods. In the regime where the detectable amplitude scales $\propto P$, a linear fit is adequate to detect the signal\footnote{It is well known, for example in pulsar timing (Joshi \& Rasio 1997) that if an orbit can be measured precisely enough, the parameters of the orbit can be determined on a timescale much less than the orbital period by very accurate measurements of orbital derivatives. For circular orbits, a cubic fit $v=a+b\,t+c\,t^2+d\,t^3$ gives a period measurement $P=2\pi\sqrt{b/6d}$ which can be inserted into equation (\ref{eq:quadK}) to find the amplitude $K$. This implies that if we could make an accurate enough measurement of the radial velocity curve, the complete orbital solution could be determined.}. However, the false alarm probability used for the F-test must reflect the number of additional trials that are carried out at high frequency using the periodogram. For example, a search for long period orbits only might adopt a FAP for the F-test of 1\%. However, if short period orbits are also searched, and the number of independent frequencies for $P<T$ is $\approx 1000$, then we should choose a FAP of $\sim 10^{-5}$.

Detection efficiency plots for long period eccentric orbits are shown in Figures \ref{fig:longecc1} and \ref{fig:longecc2}. Figure \ref{fig:longperiod} and the related discussion makes it straightforward to understand the detectability in this case. At very long periods, the number of parameters needed to describe the data is 1 or 2 as discussed above, and therefore the scalings with period are the same as the LS periodogram. However, the transition into the long period regime happens at longer orbital periods for eccentric orbits, because of the distorted nature of the lightcurve. The width of the periastron passage is $\approx (1-e)^2P$, so that equations (\ref{eq:long1}) and (\ref{eq:long2}) still apply, but with the substitution $T\rightarrow T/(1-e)^2$. The dotted curves in Figure \ref{fig:longecc2} show the analytic formula with this substitution. Therefore at long periods, eccentric orbits are more easily detected for a given $K$.


\section{Summary and Discussion}
\label{sec:conclusion}

The main results of this paper are (i) a method for estimating the significance of a Keplerian fit to radial velocity data, and (ii) an analytic expression for the velocity threshold of single planets in terms of the number and duration of the observations, number of independent frequencies, and the required false alarm probability. For circular orbits, equation (\ref{eq:sn1}) (or eq.~[\ref{eq:sn}] for $N\gg 1$) gives the velocity amplitude threshold at short periods ($P<T$). Equations (\ref{eq:long1}) and (\ref{eq:long2}) give the amplitude threshold at long periods ($P>T$) as a function of the required detection efficiency. At long periods, the 99\% detection threshold scales as $K\propto P^2\propto a^3$, whereas the 50\% detection threshold scales as $K\propto P\propto a^{3/2}$.

\begin{figure}
\epsscale{1.0}\plotone{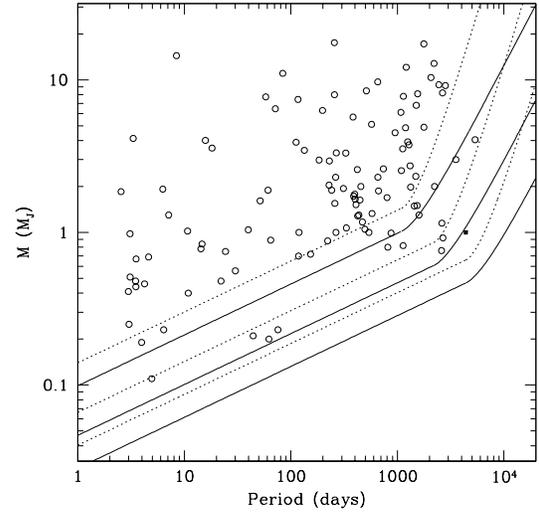}
\caption{Known companions (circles) and detection thresholds. 99\% detection thresholds are dotted lines; 50\% detection thresholds are the solid lines. In each case, we make 5 observations per year for 3, 6, and 12 years ($N=15, 30,$ and $60$), and take $\sigma=5\ {\rm m/s}$ and $M_\star=1\ M_\odot$. From top to bottom, the short period amplitude thresholds are $K_0=28, 20, 13, 9.5, 8.2,$ and $5.8\ {\rm m/s}$. We assume $M=(T/\ {\rm days})$. The solid square shows the location of Jupiter.\label{fig:mp}}
\end{figure}

We presented a straightforward generalization of the Lomb-Scargle periodogram to Keplerian orbits, based on the improvement of $\chi^2$ when a Keplerian orbit is included in the fit, and discussed a Monte Carlo method to calculate the false alarm probability associated with the fit. An simple analytic estimate of the false alarm probability is as follows. Given $\chi^2_{\rm Kep}$ from the Keplerian fit, and $\chi^2_{\rm mean}$ from the fit of a constant to the data, first calculate the power $z_0$ using equation (\ref{eq:ze}). An estimate of the false alarm probability is then ${\rm FAP}\approx M{\rm Prob}(z>z_0)$, where $M\approx T\Delta f$ is the number of independent frequencies ($\Delta f$ is the frequency range searched), and ${\rm Prob}(z>z_0)$ is the probability distribution given by equation (\ref{eq:keplerprob}). Alternatively, Figure \ref{fig:fap} may be used to find the FAP for a given $N$ and $\chi^2_{\rm mean}/\chi^2_{\rm Kep}$ ratio. This Figure should prove useful for a quick estimate of the FAP associated with a Keplerian fit.

We used the Keplerian periodogram to investigate the effect of eccentricity on detectability. Eccentricity acts to make detection more difficult at short periods, where the uneven sampling often results in inadequate phase coverage during the rapid periastron passage of an eccentric orbit. At long periods, the increased velocity amplitude and acceleration near periastron {\em increase} detectability. The transition to the long period regime occurs for orbital periods $\approx T/(1-e)^2$.

\begin{figure}
\epsscale{1.2}\plotone{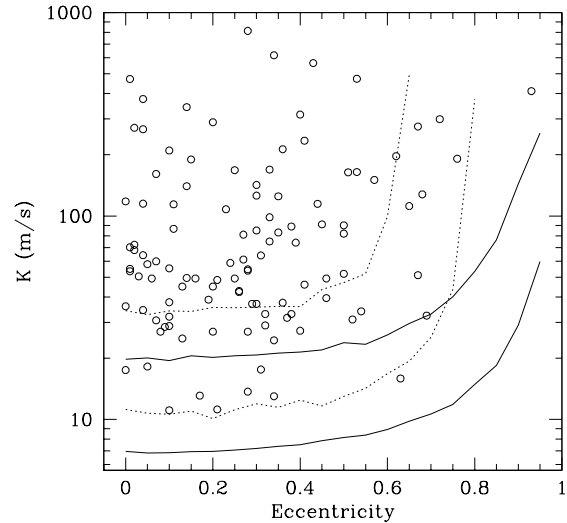}
\caption{Velocity amplitude and eccentricity of known companions. The curves show 50\% (solid) and 99\% (dotted) detection thresholds, for (top to bottom) $N=16$, and $N=39$, assuming $\sigma=5\ {\rm m/s}$.\label{fig:me}}
\end{figure}

We have also discussed the statistics of the Lomb-Scargle periodogram, including a derivation of the periodogram from basic probability theory. This Bayesian approach gives a simple derivation of the statistical distribution of periodogram powers for Gaussian noise, and clarifies the nature of different periodogram normalizations. The best statistic to use at both short and long orbital periods is $\Delta \chi^2$, the improvement in $\chi^2$ when the planet is included in the velocity fit. Using the square of the fitted amplitude $K^2$ (Nelson \& Angel 1998) results in decreased sensitivity at long periods. A seperate analysis of $K$ and phase $\phi$ recovers this sensitivity at long periods (Eisner \& Kulkarni 2001), but is unnecessary if $\Delta \chi^2$ is adopted as the statistic. 

Both Nelson \& Angel (1998) and Eisner \& Kulkarni (2001) argue for the superiority of a ``least-squares'' rather than ``periodogram'' approach to this problem. Partly, this is based on a preference for dealing directly with the parameters of the fit (amplitude $K$ and phase $\phi$ for circular orbits) and the resulting $\chi^2$, rather than a ``black box'' periodogram. For example, the original LS periodogram must be ``modified'' to include a constant term as a free parameter at each frequency (Walker et al.~1995; Nelson \& Angel 1998; CMB99), whereas this arises naturally when thinking about a $\chi^2$ fit of a model to the data. In addition, since the form of the signal is exactly known, i.e.~a Keplerian orbit, the argument is that Fourier or spectral analysis of the data is not the most efficient way to look for the signal. We hope in this paper to have clarified the equivalence of the least squares and periodogram approaches when $\Delta\chi^2$ is used as the test statistic.

We have considered only single planets in this paper, whereas multiple systems of planets are common. In cases where planets are well-seperated in period or in amplitude, our results may be applicable. For example, a linear trend in the velocities is often included in orbit fits to subtract any long term velocity variations due to a long period companion. Our results apply to this case if the number of degrees of freedom is reduced by 1, so that $N-1$ becomes $N-2$, $N-3$ becomes $N-4$ etc.~(see Walker et al.~1995; CMB99 for discussion of linear and quadratic ``background models''). Of course, this correction is only relevant for small $N$, for large $N$, the results carry over directly. Another case in which two planets are close in period, but well-seperated in amplitude is discussed by Narayan, Cumming, \& Lin (2004, in preparation).

In many situations when looking for a periodic signal, it is possible to detect signals with amplitudes much less than the background noise level. In the case of planet searches, this is not the case, because of two factors. The first, which we have discussed in this paper, is statistical. The small number of observations limit the detectable amplitude to $\approx 2$--$4\ \sigma$ for $N\approx 20$--$30$ (e.g.~see Figure 1). Here, $\sigma$ refers to a combination of measurement errors, both statistical and systematic, and intrinsic stellar ``jitter''. When the number of observations is less than $\sim 10$, it is impossible to characterize an orbit, as was previously pointed out by Nelson \& Angel (1998). Detection of signals $<1\ \sigma$ requires $N\gtrsim 50$. 

The second factor which limits detectability is uncertainty surrounding the stellar jitter. This may arise from convective inhomogeneities, or rotational modulation of magnetic features on the surface (Saar \& Donahue 1997), all processes with characteristic timescales comparable to extrasolar planet orbital periods. Therefore, although the magnitude of the stellar jitter can be estimated based on stellar properties (Saar et al.~1998), its time variability is a significant source of uncertainty. Planet detections with $K<\sigma$ require a much better understanding of jitter. Observations of magnetic activity indicators simultaneous with the radial velocity measurements offer some hope of correcting for these effects (Saar \& Fischer 2000; Paulson et al.~2002),  but this work is in its early stages. Improvement in measurement errors to the $\sim 1\ {\rm m/s}$ level will help to disentangle systematic errors and stellar jitter effects.

Understanding the distribution of planet orbital period, mass, and eccentricity at low masses, long orbital periods, and large eccentricities requires careful analysis of the radial velocity data for each survey (e.g.~Walker et al.~1995; Cumming et al.~1999; Cumming et al.~2003; Santos et al.~2003). Nonetheless, it is interesting to compare the detection thresholds we find in this paper with the observed planet properties. For a mass $M_P$, the velocity amplitude is
\begin{equation}\label{eq:K}
K={28.4\ \mathrm{m/s}\over\sqrt{1-e^2}}\ \left({M_P\sin i\over M_J}\right)\left({P\over 1\ \mathrm{yr}}\right)^{-1/3}\left({M_\star\over M_\odot}\right)^{-2/3},
\end{equation}
where $P$ is the orbital period, $M_P$ is the mass of the planet, and $M_\star$ is the mass of the star. In Figures \ref{fig:mp} and \ref{fig:me}, we show the mass, periods, and eccentricities of known planets compared to the detection curves. In Figure \ref{fig:mp}, we show curves of 50\% and 99\% detection efficiency for $\sigma=5\ {\rm m/s}$ and 3, 6, and 12 years of observations with 5 observations per year. These curves roughly match the observed cutoffs at low masses and long periods. Jupiter's position is indicated by a black square for comparison, indicating that Jupiter analogs will be detectable in the near future. In Figure \ref{fig:me}, we show 50\% and 99\% detection efficiency curves for short period orbits, and compare them with the observed $K$-$e$ distribution. The lack of observed highly eccentric orbits is possibly due to physical effects which limit their survival (e.g.~Ford, Havlikova, \& Rasio 2001). However, {\em our results emphasise that there remain significant selection effects against eccentric orbits for $e\gtrsim 0.6$, and the small number of highly eccentric planets discovered so far may reflect this}. In this regard, it is worth noting that the most eccentric orbit discovered so far, with $e=0.93$ (HD~80606; Naef et al.~2001), also has one of  the largest amplitudes $K=411\ {\rm m/s}$ ($M\sin i=4\ M_J$). Future observations may reveal lower mass planets in highly eccentric orbits that have hereto gone undiscovered.

\section*{Acknowledgements}

I thank Geoff Marcy and Greg Laughlin for useful discussions and insights, Alex McDaniel for comments on an earlier version of the manuscript, and Ira Wasserman, Daniel Reichart, and Greg Ushomirsky for useful suggestions and conversations about Bayesian statistics. The referee kindly pointed me towards references in the statistics literature. I acknowledge support from NASA through Hubble Fellowship grant HF-01138 awarded by the Space Telescope Science Institute, which is operated by the Association of Universities for Research in Astronomy, Inc., for NASA, under contract NAS 5-26555.

\end{document}